\documentclass[pra,twocolumn,preprintnumbers,amsmath,amssymb]{revtex4}
\usepackage{amsmath}
\usepackage{hyperref}

\begin{document}

\title{Quantum field theory as a bilocal statistical field theory}
\author{S.~Floerchinger}
\affiliation{Institut f\"{u}r Theoretische Physik\\Universit\"at Heidelberg\\Philosophenweg 16, D-69120 Heidelberg}
\email{s.floerchinger@thphys.uni-heidelberg.de}

\begin{abstract}
We develop a reformulation of the functional integral for bosons in terms of bilocal fields. Correlation functions correspond to quantum probabilities instead of probability amplitudes. Discrete and continuous global symmetries can be treated similar to the usual formalism. Situations where the formalism can be interpreted in terms of a statistical field theory in Minkowski space are characterized by violations of unitarity at very large momentum scales. Renormalization group equations suggest that unitarity can be essentially restored by strong fluctuation effects.
\end{abstract}

%\pacs{}

\maketitle

\section{Introduction}
The question whether the fundamental constituents of nature are particles or waves is discussed since a long time. The central equation of nonrelativistic quantum mechanics -- Schr\"{o}dingers equation \cite{Schroedinger} -- is an equation for a wave function $\psi(\vec x,t)$. On the other side, as was first pointed out by Born \cite{Born}, this wave function, or more precisely its modulus square $|\psi(\vec x,t)|^2$, can be interpreted as the probability density to detect a particle at the point $\vec x$ at time $t$. Indeed, the signals found with experimental devises such as fog chambers or silicon detectors are localized tracks as expected for pointlike particles. In this sense an electron for example has properties of both a particle and a field. An alternative view is that they always behave like waves and that the particle-like properties arise from the special case of a wavefunction that is localized in position space, i.~e. $\psi(\vec x, t)\sim \delta^{(3)}(\vec x-\vec x^\prime(t))$. However, such localized states are usually destroyed by the unitary time evolution corresponding to Schr\"odingers equation. To explain the experimentally found particle track one has to assume a frequent ``collapse of the wavefunction'' occurring in the measurement process. The time evolution during the measurement is therefore not described by a unitary evolution operator, in contrast to situations where no measurement is performed. Moreover, the fact that this collapse of the wavefunction happens instantaneously has lead to interpretational problems expressed in terms of different ``paradoxa'', most prominently the one of Einstein, Podolski and Rosen \cite{EPR}.

In our modern understanding quantum mechanics is the non-relativistic limit of the underlying quantum theory of fields. In this framework particles are seen as ``excitations of the vacuum''. However, this notion is not very precise and seems to be motivated mainly by perturbation theory for vanishing or small interactions where creation and annihilation operators for particles are introduced similar to the creation and annihilation operators for excitations of the harmonic oscillator in quantum mechanics. Concerning the interpretational difficulties mentioned above, quantum field theory does not lead to much new insight. The formulation of the theory is mainly in terms of wave functions or fields, but one still needs the interpretation in terms of probabilities to ``find a particle in a certain state'' to bring the formalism into contact with experiments. This includes also the collapse of the wavefunction connected with all the conceptual difficulties. For a recent, clear written discussion of these issues see \cite{Nikolic}. 

Despite some problems with interpretation or foundational issues, quantum field theory is a very successful theory in praxis. It describes a wide range of physical phenomena ranging from the Standard model of elementary particle physics to effective theories at low energies and momenta. The quantitative agreement between some theoretical predictions and experimental observation is amazing. It is therefore sensible to investigate foundational questions concerning the nature of particles (or fields) in this framework.

A modern formulation of quantum field theory that is not restricted to perturbation theory is the functional integral approach \cite{Feynman}. This formalism is quite close to formulations of statistical field theory \cite{ZinnJustinWeinberg}, \cite{Wetterich:2002fy}. The far reaching parallels in both formalisms were important for many theoretical achievements, for example the development of the renormalization group theory \cite{Wilson}, \cite{WegnerHoughton}. One might ask whether this duality between quantum and statistical field theory has a deeper physical origin and whether it can help us for a better understanding of foundational questions and of what ``particles'' are. One might hope that some difficult points in the interpretation of quantum mechanics related to the ``collapse of the wave function'' could be clarified.

On first sight the functional integral formulation of quantum field theory has the disadvantage, that the concept of particles does not appear explicitly. In contrast to the canonical or operator formalism, no creation and annihilation operators appear. In fact, the formulation is in terms of a weighted sum over field configurations, expectation values, correlation functions, fluctuations and so on. It is a priori not clear how particle-like properties are embedded in this framework. To relate the formalism to experiments one has to go an indirect way. The results calculated in the functional integral formalism must be interpreted by translating them to the operator language. In this way one can for example extract information about the scattering matrix that appears in the canonical formalism from the correlation functions calculated with the functional integral. Although this works in praxis, it would be desirable to have a closed interpretation of the functional integral for conceptual reasons. This approach has many advantages, for example it is manifestly Lorentz-invariant, can be applied also for gauge theories and strong interactions and captures non-perturbative features. 

The fact that particles are somehow hidden in the functional integral formulation can also be seen as a chance. After all, also the connection of the canonical formalism to experiments needs nontrivial interpretational steps. One has to interpret the modulus square of some wavefunction as a probability density to ``detect a particle'' and the formalism includes the collapse of the wavefunction with all its difficulties. One might hope, that a direct interpretation of the functional integral -- if it can be established -- does not suffer from these problems. One reason is that the formalism is completely Lorentz-invariant. This also being true on the interpretational level could solve some problems coming from the fact that the collapse of the wavefunction by being instantaneous breaks Lorentz-invariance.

In the canonical formalism the physics of $n$ (conserved) particles is described by a $n$-particle state in Fock space and its time evolution. Fock states do not have a direct correspondence in the functional integral formalism. Here, the same physics is described by a $2n$-point correlation function. Instead of describing the state at a given time, these correlation functions address directly the evolution and have two space-time arguments for every particle -- typically one corresponding to the initial and the other to the final time. At the first rough glance this seems to be already close to the experimental situation where indeed some sort of correlation between initial ``measurements'' (or preparation of state) and final measurements are considered. However, one must keep in mind that the correlation functions used in the functional integral description are not quite the ones known from statistical theories. Both meaning and properties differ. 

In statistical field theory one calculates expectation values, correlation functions etc. by taking the sum over all possible field configurations $\varphi$ weighted by the real and positive semi-definite probability measure
\begin{equation}
p[\varphi]=e^{-S[\varphi]},
\end{equation}
which we call ``functional probability''. For a local theory in $d$ dimensions the action $S[\varphi]$ is given as an integral over the Lagrange density
\begin{equation}
S[\varphi]=\int d^d x {\cal L}
\end{equation}
where ${\cal L}$ is a function of $\varphi$ and its derivatives. In quantum field theory, the corresponding measure under the functional integral
\begin{equation}
e^{iS[\varphi]}
\end{equation}
is complex and can therefore not be interpreted as a probability. In addition, to calculate transition probabilities from the correlation functions one has to take the modulus square at some point. Despite this subtleties, one can transfer a lot of knowledge and intuition from statistical to quantum field theories. 

We propagate here a scenario where the fundamental constituents of nature are bilocal microscopic quantum fields. In the language of Bell, they are the ``beables'' \cite{BellBeables}, i.~e. they exist in the sense of an objective reality. However, this does not mean, that this microscopic fields can be measured directly. Only parts are accessible, namely the ones that have particle-like properties. Particles are not fundamental objects but arise on a secondary, effective or derived level. It is conceivable that a description in particle-related terms makes sense under special circumstances, only. For example, it may be that the concept of particles is useful in measurement-like situations while in other situations this is not the case.

The formulation of the field theory is in terms of probabilities for microscopic field configurations. No assumption is made whether these probabilities arise from a underlying deterministic theory. It may well be that this is not the case such that the description of nature is probabilistic already on the fundamental level.

The idea to abandon particles as the fundamental constituents of nature was also put forward by Davies \cite{Davies}. From an investigation of the particle concept in curved space-time he comes to the conclusion ``The concept of particles is purely an idealized model of some utility in flat space quantum field theory. Away from that limited context, however, the concept becomes much less useful [...]''. Many problems discussed in ref.\ \cite{Davies} might find a solution in the scenario proposed here, where the fundamental description is in terms of the microscopic fields and their correlations. 

Since in our scenario particles are not fundamental but emerging objects, one cannot assign them properties in the sense of an objective reality. Partly due to this feature, our scenario is in accordance with the theorems of Bell \cite{Bell} or Kochen and Specker \cite{KochenSpecker}. Anyhow, the approach put forward in this paper does not constitute a hidden variable theory, since the description of the fundamental entities -- the microscopic fields -- is probabilistic.

This paper is organized as follows. In section \ref{sec:BilocalFI} we review briefly the functional integral formulation of quantum field theory for bosonic fields and discuss a rewriting in terms of a bilocal field. Discrete and continuous symmetries of this approach are discussed in section \ref{sec:Symmetries} and positive correlations in section \ref{sec:positiviCorrelations}. In the subsequent section \ref{sec:flowequation} we propose a scenario that allows an interpretation of the bilocal functional integral formalism in terms of a statistical field theory for Minkowski signature. We provide arguments for this by investigating renormalization group equations for an Euclidean theory with complex action. In section \ref{sec:Propertiesbilocal} we discuss the properties of a bilocal statistical field theory and sketch how it can be used to describe concrete experiments. Finally, we draw some conclusions in section \ref{sec:conclusions}.

\section{A bilocal functional integral}
\label{sec:BilocalFI}
We start this section with a brief review of the standard functional integral for a bosonic scalar field $\varphi(t,\vec x)$. For simplicity, we assume that $\varphi$ has a vanishing vacuum expectation value for vanishing external source $J$. If this was not the case, we could easily redefine the fields according to $\varphi^\prime(x) = \varphi(x) -\varphi_\text{VEV}$. A crucial object for the functional integral formulation is the partition function
\begin{equation}
Z[J] = \int D\varphi \; e^{iS[\varphi]-i\int_x \left\{J^*(x)\varphi(x)+\varphi^*(x) J(x)\right\}}.
\label{eq:nu0}
\end{equation}
For a local quantum field theory, the action $S$ is given by
\begin{equation}
S[\varphi] = \int_x {\cal L},
\label{eq:}
\end{equation}
where ${\cal L}$ is a Lagrange density that depends on the field $\varphi$ and its derivatives at the point $x$. The space-time coordinate $x=(x_0,\vec x)$ contains both the time $x_0$ and the spatial position $\vec x$ and $\int_x$ is an abbreviation for $\int dx_0 \int d^3x$. Form the partition function we can obtain the source-dependent expectation value according to
\begin{eqnarray}
\nonumber
\langle \varphi(x) \rangle &=& \frac{1}{Z[J]} \int D\varphi \; \varphi(x) e^{iS[\varphi]-i\int_x \{J^*\varphi + \varphi^*J\}}\\
&=& i \frac{1}{Z[J]} \frac{\delta}{\delta J^*(x)} Z[J],
\label{eq:}
\end{eqnarray}
and the assumption made above implies that this vanishes for $J=0$. It is useful to introduce the functional $W[J]$ according to
\begin{equation}
Z[J] = e^{-iW[J]}
\end{equation}
such that
\begin{equation}
\langle \varphi(x) \rangle = \frac{\delta}{\delta J^*(x)} W[J].
\label{eq:}
\end{equation}
Another important object is the two-point correlation function for $J=0$
\begin{equation}
D_F(x,y) = \langle \varphi(x) \varphi^*(y) \rangle
\label{eq:}
\end{equation}
which is obtained from the partition function as
\begin{eqnarray}
\nonumber
D_F(x,y) &=& i^2 \frac{1}{Z[0]} \frac{\delta}{\delta J^*(x)} \frac{\delta}{\delta J(y)} Z[J]{\big |}_{J=0},\\
&=& i \frac{\delta}{\delta J^*(x)} \frac{\delta}{\delta J(y)} W[J]{\big |}_{J=0}.
\label{eq:}
\end{eqnarray}
In the last line we used that $\delta W/ \delta J$ vanishes for $J=0$. 

The two-point function can be related directly to a quantity that is accessible to experiments. For a single particle that has been detected at the space-time point $y$, the probability to find it at the position $\vec x$ at time $x_0>y_0$ is proportional to
\begin{equation}
p(x,y) = |D(x,y)|^2 = (\langle \varphi(x) \varphi^*(y) \rangle)^* \langle \varphi(x) \varphi^*(y)\rangle.
\label{eq:}
\end{equation}
This transition probability can be expressed as a ``double'' functional integral
\begin{eqnarray}
\nonumber
p(x,y) &=& \frac{1}{Z[0]Z^*[0]} \int D\varphi_+ D \varphi_- \; \varphi_+(x) \varphi_-(x) \\
&&\times \left(\varphi_+(y) \varphi_-(y)\right)^* e^{iS[\varphi_+]} e^{-iS^*[\varphi_-]}.
\label{eq:nu1}
\end{eqnarray}
Here we work with the fields $\varphi_+$ and $\varphi_-$ which should be considered as independent. Indeed, we can easily pull the two functional integrals in Eq.\ \eqref{eq:nu1} apart, such that the terms involving $\varphi_+$ give $D_F(x,y)$ while the terms involving $\varphi_-$ give the complex conjugate expression. For usual quantum field theories the action $S$ is real up to a small term that ensures the convergence of the functional integral and is important at a later stage for the description of contour integrals. For example, the action $S$ might be of the form (we use the metric $\eta=\text{diag}(-1,1,1,1)$)
\begin{equation}
S[\varphi] =  \int_x \varphi^*(x) (\partial_\mu \partial^\mu -m^2+i\epsilon) \varphi(x) + S_\text{int}.
\label{eq:ad1}
\end{equation}
The conjugate action $S^*[\varphi]$ is then of the same form with $i\epsilon$ replaced by $-i\epsilon$. More generally it is defined by $(S[\varphi])^*=S^*[\varphi^*]$.

We can obtain the transition probability in Eq.\ \eqref{eq:nu1} also from the generalized partition function
\begin{eqnarray}
\label{eq:nu2}
\nonumber
Z[J_+, J_-]  &=&  \int D \varphi_+ D \varphi_-\; \exp{\bigg (} i S[\varphi_+]-iS^*[\varphi_-]\\
\nonumber
&& -i\int_x \{J_+^*\varphi_++c.c.\}+i \int_x \{J^*_-\varphi_-+c.c.\}{\bigg )}.\\
\end{eqnarray}
We note that Eq.\ \eqref{eq:nu2} contains the functional integral over the field configurations $\varphi_+$ and $\varphi_-$. One may also write this as a single functional integral over the fields that depend in addition to the position variable $\vec x$ on the contour time $t_c$ which is integrated along the Keldysh contour \cite{Keldysh}. However, for our purpose it will be more convenient to work directly with the expression in Eq.\ \eqref{eq:nu2}. 

It is clear that the modulus square of correlation functions such as the transition probability $p(x,y)$ can be obtained from the partition function in Eq. \eqref{eq:nu2} by taking appropriate derivatives with respect to the sources $J_+$ and $J_-$. Since we are interested in the modulus square, derivatives with respect to $J_+$ and $J_-$ will allways appear in pairs. For example
\begin{eqnarray}
&& p(x,y) = \frac{1}{Z[J_+,J_-]}\\
\nonumber
&& \frac{\delta}{\delta J^*_+(x)} \frac{\delta}{\delta J^*_-(x)} \frac{\delta}{\delta J_+(y)} \frac{\delta}{\delta J_-(y)} Z[J_+,J_-] {\big |}_{J_+=J_-=0}.
\label{eq:}
\end{eqnarray}

Such correlation functions can also be obtained from a partition function involving bilocal source terms. It is defined as
\begin{eqnarray}
\nonumber
Z[K]  &=& \int D \varphi_+ D\varphi_- \; \exp{\bigg \{}iS[\varphi_+]-iS^*[\varphi_-]\\
&&+\int_{x,y}\{K^*(x,y)\varphi_+(x)\varphi_-(y)+ c.c. \}{\bigg \}}.
\label{eq:nu3}
\end{eqnarray}
The source field $K$ depends on the space-time variables $x$ and $y$. The transition probability is obtained from functional differentiation according to
\begin{equation}
p(x,y) = \frac{1}{Z[K]} \frac{\delta}{\delta K^*(x,x)} \frac{\delta}{\delta K(y,y)} Z[K]{\big |}_{K=0}.
\label{eq:nu3a}
\end{equation}
At this point two remarks are in order.

(1) The bilocal source term $K(x,y)$ is of bosonic nature. For bosonic fields $\varphi_+$ and $\varphi_-$ this is obvious but also when $\varphi_+$ and $\varphi_-$ are fermionic fields, it is the case. indeed, the source $K^*$ couples to the bilinear function $\varphi_+\varphi_-$ which is a bosonic operator also for Grassmann-valued fields $\varphi_+$ and $\varphi_-$.

(2) Derivatives of $Z[K]$ at $K=0$ always factorize into a functional integral over the field $\varphi_+$ and over the field $\varphi_-$. This is not the case for $K\neq 0$, however.

We can formally rewrite Eq.\ \eqref{eq:nu3} in terms of a functional integral over a bilocal field $\chi(x,y)$. For this purpose we introduce under the functional integral on the right hand side of Eq.\ \eqref{eq:nu3} a factor of $1$ in the form
\begin{equation}
1= \int D \chi \;\delta[\chi],
\label{eq:}
\end{equation}
where $\delta[\chi]$ is a functional Dirac distribution defined by
\begin{equation}
\int D \chi \;\delta[\chi] F[\chi] = F[0]
\label{eq:}
\end{equation}
for a regular functional of the bilocal field $\chi(x,y)$ and its complex conjugate. The notation is such that $\delta[\chi]$ is a distribution for both the real and imaginary part of $\chi$. An example for a concrete realization is 
\begin{equation}
\delta[\chi] = \lim_{\alpha\to\infty} c(\alpha) \exp\left(-\alpha \int_{x,y} \chi^*(x,y)\chi(x,y)\right),
\end{equation}
 where $c(\alpha)$ is an irrelevant constant.

After a shift in the integration variable $\chi$, we obtain from Eq.\ \eqref{eq:nu3}
\begin{eqnarray}
\nonumber
Z[K]  &=& \int D\varphi_+ D\varphi_- D\chi \; \delta[\chi-\varphi_+\varphi_-] \\
&&\times e^{i S[\varphi_+]-iS^*[\varphi_-]} e^{\int_{x,y}\{K^*(x,y)\chi(x,y)+c.c.\}}.
\label{eq:}
\end{eqnarray}
We can now formally perform the functional integral over $\varphi_+$ and $\varphi_-$ such that this becomes
\begin{equation}
Z[K] = \int D\chi \; e^{-Q[\chi]} \; e^{\int_{x,y} \{K^*(x,y)\chi(x,y)+\chi^*(x,y)K(x,y)\}}
\label{eq:nu4}
\end{equation}
with 
\begin{equation}
e^{-Q[\chi]} = \int D\varphi_+ D\varphi_- \; \delta[\chi-\varphi_+\varphi_-] e^{iS[\varphi_+]-iS^*[\varphi_-]}.
\label{eq:nu5}
\end{equation}
Eq.\ \eqref{eq:nu4} gives a description of our quantum field theory in terms of a bilocal bosonic field $\chi$.

\section{Symmetries}
\label{sec:Symmetries}
In the following we discuss the transformation behavior of $Q[\chi]$ under various discrete and continuous transformations of the argument $\chi$. As before we assume that $\varphi_+$ and $\varphi_-$ are bosonic scalar fields. 

For the bilocal field $\chi(x,y)$ the charge conjugate field is defined by $\chi_c(x,y)=\chi^*(x,y)$. For the bilocal action $Q[\chi]$ we find
\begin{eqnarray}
e^{-Q[\chi_c]} &=& \int D \varphi_+ D \varphi_- \; \delta[\chi-(\xi^*\varphi_+^*)(\xi\varphi^*_-)]\\
&& e^{iS[\varphi_+]} e^{-iS^*[\varphi_-]}.
\label{eq:}
\end{eqnarray}
Here $\xi$ is a complex phase ($\xi^*\xi=1$) that determines the intrinsic charge parity of the scalar field $\varphi_+$. After relabeling the integration variable $\xi\varphi_+ \to \varphi_+^*$, $\xi^*\varphi_+^* \to \varphi_+$, $\xi^*\varphi_-\to \varphi_-^*$, $\xi\varphi_-^*\to \varphi_-$ this gives
\begin{equation}
e^{-Q[\chi_c]} = \int D \varphi_+ D \varphi_- \; \delta[\chi-\varphi_+\varphi_-] e^{iS_c[\varphi_+]} e^{-i S_c^*[\varphi_-]}
\label{eq:}
\end{equation}
with $S_c[\varphi]=S[\xi^*\varphi^*]$ the charge-conjugated microscopic action. It follows that $Q[\chi_c]=Q[\chi]$ when $S$ is invariant under charge conjugation, i.e. $S[\varphi] = S_c[\varphi]$.

Very similar to charge conjugation are parity and time reversal. The transformations are defined for the bilocal field $\chi$ similar as for the local fields $\varphi_+, \varphi_-$ but as for charge conjugation no complex phase (intrinsic parity) appears for $\chi$. In general, $Q$ will  show the same symmetries as does $S$, in particular under the combined transformation CPT.

As a final discrete transformation we discuss hermitian conjugation of $\chi$, i.e. complex conjugation together with an interchange of the arguments
\begin{equation}
\chi_s(x,y)=\chi^*(y,x).
\end{equation}
This gives for the bilocal action
\begin{eqnarray}
\nonumber
e^{-Q[\chi_s]} &=& \int D\varphi_+ D\varphi_- \; \delta[\chi- \varphi_-^* \varphi_+^*] \; e^{iS[\varphi_+]} e^{-iS^*[\varphi_-]}.\\
\label{eq:defQ}
\end{eqnarray}
After relabeling the integration variables according to $\varphi_+ \to \varphi_-^*$, $\varphi^*_+ \to \varphi_-$, $\varphi_- \to \varphi_+^*$,  $\varphi_-^* \to \varphi_+$ this becomes
\begin{eqnarray}
\nonumber
e^{-Q[\chi_s]} &=& \int D \varphi_+ D\varphi_- \; \delta[\chi-\varphi_+\varphi_-]\; e^{-iS^*[\varphi_+^*]} e^{iS[\varphi_-^*]}\\
&=& (e^{-Q[\chi]})^*.
\label{eq:}
\end{eqnarray}
Note that this implies that $e^{-Q[\chi]}$ is real when restricted to hermitian bilocal fields with $\chi=\chi_s$.

We now turn to the continuous symmetries. As an example we consider a global $U(1)$ transformation
\begin{equation}
\chi(x,y)\to e^{i\alpha} \chi(x,y).
\label{eq:}
\end{equation}
The bilocal action transforms according to
\begin{eqnarray}
\nonumber
e^{-Q[e^{i\alpha}\chi]} &=& \int D\varphi_+ D\varphi_- \; \delta[\chi-e^{-i\alpha} \varphi_+ \varphi_-] \\ 
&& e^{iS[\varphi_+]} e^{-iS^*[\varphi_-]}.
\label{eq:}
\end{eqnarray}
For an action $S[\varphi]$ that is invariant under a global $U(1)$ transformation we can change the integration variable, for example according to
\begin{equation}
\varphi_+ \to e^{i\alpha} \varphi_+, \quad \varphi_+^*\to e^{-i\alpha} \varphi_+^*,
\label{eq:}
\end{equation}
which gives
\begin{equation}
Q[e^{i\alpha}\chi] = Q[\chi].
\label{eq:}
\end{equation}
In a similar way, other continuous symmetries of $S$ find their corresponding counterpart for the bilocal action $Q$.

\section{Positive correlations}
\label{sec:positiviCorrelations}
The fact that $Q$ can be written in terms of a double functional integral as in Eq.\ \eqref{eq:nu5} has some interesting consequences. Although the weight $e^{-Q[\chi]}$ in Eq.\ \eqref{eq:nu4} is in general complex, some expectation values and correlation functions are real and positive. Consider for example the transition probability in Eq.\ \eqref{eq:nu3a},
\begin{equation}
p(x,y) = \frac{1}{Z[K=0]} \int D \chi \; \chi(x,x) \chi^*(y,y) e^{-Q[\chi]}.
\label{eq:nu6}
\end{equation}
Since this can be written as a squared amplitude correlation function as in Eq.\ \eqref{eq:nu1} it is a real and positive semidefinite function of the arguments $x$ and $y$. From the expression in Eq.\ \eqref{eq:nu6} this is not obvious but must be seen as a property of the bilocal action $Q[\chi]$. 

Similar to the transition probability in position space one can construct other positive correlation functions. Let us consider a complete orthonormal set of functions $f_n(x)$ with the properties
\begin{eqnarray}
\nonumber
\int_x f_n^*(x) f_m(x) &=& \delta_{nm},\\
\sum_n f_n^*(x) f_n(y) &=& \delta(x-y),
\label{eq:}
\end{eqnarray}
where $\delta(x-y)=\delta(x_0-y_0) \delta^{(3)}(\vec x-\vec y)$. We define the bilocal fields in this basis by
\begin{equation}
\chi(n,m) = \int_{x,y} f_n^*(x) f_m(y) \chi(x,y),
\end{equation}
such that
\begin{equation}
\chi(x,y) = \sum_{n,m} \chi(n,m) f_n(x) f^*_m(y).
\label{eq:}
\end{equation}
Note that $\chi(n,m)$ is hermitian (antihermitian) with respect to the arguments $n$ and $m$ if and only if $\chi(x,y)$ is hermitian (antihermitian). 

Correlation functions that involve the modes $\chi(n,m)$ with equal indices $n=m$ will be real and positive semi-definite, for example
\begin{equation}
\frac{1}{Z[K=0]} \int D \chi \, \chi(n,n) \chi^*(\tilde m,\tilde m) e^{-Q[\chi]}\geq 0.
\label{eq:nd1}
\end{equation}
This is immediately clear since these functions can be written as a product of an amplitude correlation function with its complex conjugate. Moreover, we can choose a different basis for the field which is labeled with the index $\tilde m$, i.e.
\begin{equation}
\chi(\tilde n,\tilde m) = \int_{x,y} g_{\tilde n}^*(x) g_{\tilde m}(y) \chi(x,y)
\label{eq:}
\end{equation}
(and similar for $\chi^*(\tilde n, \tilde m)$) with an orthonormal set of functions $g_{\tilde n}$. We see that the modes of $\chi$ that can be written as $\chi(n,n)$ in some basis have the special and interesting property that expectation values and correlation functions are positive semi-definite. In the following, these modes are therefore called ``positive correlated modes''. Of special interest for experiments are such sets of functions that are ``local in time'', i.e. of the form
\begin{equation}
f_{(t,n)}(x) = \delta(x_0-t) \phi_n(\vec x).
\label{eq:lo2}
\end{equation}
We combined here the time $t$ and the index $n$ to a combined index $(t,n)$. The bilocal field in this basis reads $\chi((t_1,n_1),(t_2,n_2))$ and expectation values such as
\begin{equation}
\frac{1}{Z}\int D\chi \, \chi((t_2,n_2),(t_2,n_2)) \; \chi^*((t_1,n_1),(t_1,n_1)) e^{-Q[\chi]}
\label{eq:lo3}
\end{equation}
describe the probability for a particle that was prepared in the state labeled by $n_1$ at time $t_1$ to be found in the state corresponding to the index $n_2$ at time $t_2$. From the equivalence of this to a squared amplitude correlation function it is clear that this is a real and positive semi-definite function.

We note that a basis of the form in Eq.\ \eqref{eq:lo2} has an interesting feature in situations with translational invariance in time direction of the action $S[\varphi]$. In that case correlation functions as in Eq.\ \eqref{eq:lo3} are invariant with respect to an overall shift in time but also with respect to a transformation where only the first time argument of every bilocal field is shifted, e.g.
\begin{equation}
 \chi((t,n),(t,n))\to\chi((t+\Delta,n),(t,n)).
\end{equation}
This is also clear from the prescription in terms of a squared amplitude correlation function.

To close this section we remark that the positivity property as in Eq.\ \eqref{eq:nd1} is independent of the microscopic action $S[\varphi]$ that goes into the definition of the bilocal action in Eq.\ \eqref{eq:nu5}. It may be a real functional but can contain also an imaginary part of arbitrary size.

\section{Probabilistic interpretation and renormalization group flow}
\label{sec:flowequation}
So far, the formalism developed in the previous sections gives only a reformulation of ordinary quantum field theory. Similar to the action $S[\varphi]$, there is no probabilistic interpretation of $Q[\chi]$ since it is in general a complex functional and $e^{-Q[\chi]}$ cannot be seen as a functional probability. (For hermitian $\chi$ one has $e^{-Q[\chi]}\in {\mathbb R}$.) We will now examine the meaning of $Q[\chi]$ more closely and propose a scenario that leads to real and positive functional probabilities $e^{-Q[\chi]}$. 

The bilocal action $Q[\chi]$ is hard to calculate. Its definition in Eq.\ \eqref{eq:nu5} shows that it is in a non-trivial relation to the action $S[\varphi]$ even when the latter is of a Gaussian form. This makes it hard to develop an intuition or to calculate $Q[\chi]$ even approximately. The situation is better for the partition function $Z[K]$ or the correlation functions that follow as functional derivatives of this object. The correlation functions that can be obtained from $Z[K]$ by taking derivatives with respect to $K$ and setting $K=0$ subsequently can also be obtained as a product of two amplitude correlation functions. These in turn can be calculated exactly for Gaussian actions $S[\varphi]$ and many approximation methods exist for theories with interactions.

Often, the correlation functions are directly related to objects that can be measured in experiments such as scattering cross-sections or decay rates. While correlation functions are accessible to experiments,  this is different for the action $S[\varphi]$ itself. In principle, the complete knowledge of all (independent) correlation functions would allow the determination of $S[\varphi]$, but in praxis this is not helpful. With low energy experiments one can only probe an effective action $\Gamma[\varphi]$ where -- loosely speaking -- high energy fluctuations have been integrated out already. The determination of $S[\varphi]$ itself is not possible. A connection between the effective action $\Gamma[\varphi]$ and the microscopic action $S[\varphi]$ is given by renormalization group flow equations. Due to the nonlinear structure of these equations, it is in praxis not possible to determine $S[\varphi]$ uniquely by starting from the low energy action $\Gamma[\varphi]$ and following the flow to the ultraviolet.

A central question in this context is: Could the microscopic action $S[\varphi]$ be of such a form that the exponentiated bilocal action $e^{-Q[\chi]}$ is real and positive? In that case we would have a direct probability interpretation of the bilocal functional integral in Eq.\ \eqref{eq:nu4} and quantum field theory could be seen as a bilocal statistical field theory.

One possibility for this scenario is an imaginary microscopic action 
\begin{equation}
S[\varphi] = i \tilde S[\varphi],
\label{eq:nu8}
\end{equation}
with a real function $\tilde S[\varphi]$. From Eq.\ \eqref{eq:nu5} it follows that $e^{-Q[\chi]}$ would be positive semidefinite since it is a functional integral with a real and positive weight \footnote{In realistic situations $\tilde S[\varphi]$ might include an infinitesimal imaginary term $i\epsilon$ as in Eq.\ \eqref{eq:ad1} which enforces the appropriate frequency integration contour.}. Is it possible that the microscopic action $S[\varphi]$ is of the form in Eq.\ \eqref{eq:nu8} without changing the effective action $\Gamma[\varphi]$ that can be measured experimentally? We will discuss this question in the remainder of this section. We note however, that the form in Eq.\ \eqref{eq:nu8} is not an necessary condition for a real and positive functional probability $e^{-Q[\chi]}$. For example, $S[\varphi]$ can always contain a constant ($\varphi$-independent) real part which would change the overall normalization of $e^{-Q[\chi]}$, only.

Although we are ultimately interested in quantum field theories with Minkowski signature of space-time, it is often useful to study theories with Euclidean signature. The reason is that both types of theories can be connected by analytic continuation. As an example we consider a theory for a bosonic scalar field $\varphi$ with action as in Eq.\ \eqref{eq:ad1}. In momentum space the quadratic part reads
\begin{equation}
\int_p \varphi^*(p) \left(p_0^2-\vec p^2-m^2+i\epsilon\right)\varphi(p).
\label{eq:}
\end{equation}
The $i\epsilon$ term is a reminder that we should use the Feynman prescription to evaluate loop integrals that contain the propagator corresponding to the above action. The poles for frequency integration are shifted slightly away from the real axis
\begin{equation}
p_0=\pm (\sqrt{\vec p^2+m^2}-i\epsilon).
\label{eq:}
\end{equation}
Equivalently one can drop the $i\epsilon$ term and make a rotation of the frequency integration contour according to
\begin{equation}
 p_0=e^{i\alpha} p_4
\end{equation}
where $p_4$ is now integrated from $-\infty$ to $\infty$. The time is replaced by the variable $\tau$ according to
\begin{equation}
 t=e^{-i\alpha} \tau.
\end{equation}
The choice $\alpha\to 0_+$ corresponds to Minkowski signature and $\alpha\to \pi/2$ to Euclidean signature. For the later choice the action in Eq.\ \eqref{eq:ad1} becomes
\begin{eqnarray}
\nonumber
S\to i S_E &=&i \int_{x_E} \varphi^*(x_E) \left(-\partial_\tau^2-\vec \nabla^2+m^2\right) \varphi(x_E)\\
&& + i S_{E,\text{int}},
\label{eq:}
\end{eqnarray}
with $x_E=(\tau,\vec x)$. The analytic continuation has two important features. First, the inverse microscopic propagator is now of a form where frequency and spatial momentum appear in a unified way and with equal sign. Second, the functional integral with oscillating weight
\begin{equation}
e^{iS[\varphi]}
\label{eq:}
\end{equation}
becomes now a functional integral with the real and positive weight
\begin{equation}
e^{-S_E[\varphi]}.
\label{eq:}
\end{equation}
The Euclidean action $S_E$ is obtained by replacing $t$ by $-i\tau$ at various places. It conventional to add one additional overall sign, as well. 

Note that a functional integral with real and positive weight $e^{iS[\varphi]}=e^{-\tilde S[\varphi]}$ in Minkowski space becomes after analytic continuation to the Euclidean space a functional integral over the complex and oscillating weight 
\begin{equation}
e^{-S_E[\varphi]} = e^{-i\tilde S_E[\varphi]}.
\label{eq:nu8a}
\end{equation}
To summarize, an complex oscillating weight functional with Minkowski signature corresponds to a real and positive weight functional with Euclidean signature, while a real and positive weight functional with Minkowski signature corresponds to a complex oscillating weight functional with Euclidean signature.

Of course, it is known from many experiments that the actual world has Minkowski signature. On the other side, the investigation of field theories in a functional integral is simpler with Euclidean signature. For example, the construction of regulator functions is straightforward in the later case and exact renormalization group equations can be used directly. 

Let us consider the flowing action $\Gamma_k[\varphi]$. It is defined as the effective action in presence of an infrared cutoff term
\begin{equation}
\Delta S_k[\varphi] = \int_p \varphi^*(p) R_k(p^2) \varphi(p)
\label{eq:cutoff}
\end{equation}
in the microscopic action. More precisely, we modify the Euclidean version of the functional integral in Eq.\ \eqref{eq:nu0} according to
\begin{equation}
Z_k[J] = e^{W_k[J]} = \int D\tilde \varphi\; e^{-S_E[\tilde \varphi]-\Delta S_k[\tilde \varphi]+\int_x \{J^*\tilde \varphi+c.c.\}},
\label{eq:}
\end{equation}
with $\Delta S_k$ as in Eq.\ \eqref{eq:cutoff}. Consider now the Legendre transform of $W_k[J]$
\begin{equation}
\tilde \Gamma_k[\varphi] = \int_x \{J^* \varphi +c.c.\} - W_k[J],
\label{eq:}
\end{equation}
with $\varphi=\delta W_k[J] / \delta J$. The effective average action or flowing action is defined by subtracting from this the cutoff term
\begin{equation}
\Gamma_k[\varphi] = \tilde \Gamma_k[\varphi] - \Delta S_k[\varphi].
\label{eq:}
\end{equation}

The flowing action interpolates between the microscopic action
\begin{equation}
\lim_{k\to\infty} \Gamma_k[\varphi] =S[\varphi]
\label{eq:}
\end{equation}
for large values of the infrared cutoff parameter $k$ and the quantum effective action
\begin{equation}
\lim_{k\to0} \Gamma_k[\varphi] =\Gamma[\varphi]
\label{eq:}
\end{equation}
for vanishing infrared parameter $k$. We used here the properties of the cutoff function $R_k(q^2)\to 0$ for $k\to 0$ and $R_k(q^2)\to \infty$ for $k\to \infty$. The quantum effective action is the generating functional of one-particle irreducible Feynman diagrams. For nonzero value of $k$, the flowing action $\Gamma_k[\varphi]$ is closely connected to an action for averages of fields \cite{Wetterich:Impovedaverageaction}.

With these definitions, the flowing action obeys the exact functional differential equation \cite{Wetterich:1992yh}
\begin{equation}
\partial_k \Gamma_k[\varphi] = \frac{1}{2} \text{Tr} (\Gamma_k^{(2)}[\varphi]+R_k)^{-1} \partial_k R_k.
\label{eq:nu11}
\end{equation}
The trace operation $\text{Tr}$ includes an integral over momenta and a sum over possible internal degrees of freedom. We note that the right hand side of Eq.\ \eqref{eq:nu11} contains $\Gamma_k^{(2)}$, the second functional derivative of $\Gamma_k[\varphi]$
\begin{equation}
 \Gamma_k^{(2)}[\varphi](q,q^\prime) = \frac{\delta^2\Gamma_k[\varphi]}{\delta \varphi^*(q)\delta\varphi(q^\prime)}.
\end{equation}
Since both sides have a functional dependence on $\varphi$, Eq. \eqref{eq:nu11} is a functional differential equation. For an review of the flow equation and the concept of the flowing action we refer to ref. \cite{FRGReviews}

We will now investigate the renormalization flow of an action that is rotated from the real axis to the complex plane. In the microscopic regime for large $k$ it is of the form
\begin{equation}
\Gamma_k[\varphi] = e^{i\alpha_k} \hat \Gamma_k[\varphi]
\label{eq:nu12}
\end{equation}
with $\hat \Gamma_k[\varphi]$ a real Euclidean functional of the field $\varphi$. We choose the parameterization such that $\alpha_k=0$ corresponds to the usual sign convention for an action with Euclidean signature. It is convenient to choose the cutoff function according to
\begin{equation}
\Delta S_k = e^{i\alpha_k} \int_p \varphi^*(p) \hat R_k(p^2) \varphi(p)
\label{eq:}
\end{equation}
with real and positive $\hat R_k(p^2)$. The flow equation in Eq.\ \eqref{eq:nu11} becomes now
\begin{eqnarray}
\nonumber
&&e^{i\alpha_k} (\partial_k \hat \Gamma_k+ i (\partial_k \alpha_k) \hat \Gamma_k) \\
&& = \frac{1}{2} \text{Tr} (\hat \Gamma_k^{(2)}+\hat R_k)^{-1} (\partial_k \hat R_k + i(\partial_k \alpha_k) \hat R_k).
\label{eq:nu13}
\end{eqnarray}
This should be seen as a complex equation that determines both $\partial_k \hat \Gamma_k$ and $\partial_k \alpha_k$. 

In general, the simple form in Eq.\ \eqref{eq:nu12}  will not be conserved by the flow. In an expansion of the flowing action
\begin{equation}
\Gamma_k[\phi] = \sum_n \gamma_k^{(n)} {\cal O}^{(n)}[\varphi]
\label{eq:nu14}
\end{equation}
with real operators ${\cal O}^{(n)}[\varphi]$, the coefficients $\gamma_k^{(n)}$ will be complex functions of the scale parameter $k$. In principle one can write $\gamma_k^{(n)} = e^{i\alpha_k^{(n)}} \hat \gamma_k^{(n)}$ with $\hat \gamma_k^{(n)}\in {\mathbb R}$ and investigate the flow equations for the set of phases $\alpha_k^{(n)}$. For simplicity we assume here that all these phases are equal, $\alpha_k^{(n)} = \alpha_k$, so that we can use Eq.\ \eqref{eq:nu13}. This gives useful qualitative insights. For concrete models one can of course also study truncations with a number of independent complex coefficients $\gamma_k^{(n)}$. 

We note that both sides of equation \eqref{eq:nu13} are real for $\alpha_k=0$ such that this corresponds to a renormalization group fixed point. Moreover, there can be only fixed points for $\alpha_k$ when $e^{i\alpha_k}$ is real, since the right hand side of Eq.\ \eqref{eq:nu13} is real for $\partial_t \alpha_k=0$. In particular, the flow always drives the action away from the point with $\alpha_k=\pi/2$ corresponding to Eq.\ \eqref{eq:nu8a}. In situations with strong interaction effects, the renormalization group modifications are large. In this case, the action $\Gamma_k[\phi]$ or more concrete the expansion coefficients $\gamma_k^{(n)}$ in Eq.\ \eqref{eq:nu14} will be driven towards the real axis. 

These arguments show that $\alpha_k=0$ is likely to be an infrared attractive renormalization group fixed point for a large class of initial microscopic actions $S[\varphi]$. In situations with strong running due to large fluctuation effects, this fixed point is approached very quickly. By this mechanism, renormalization group modifications could render a theory with complex or purely imaginary action $S_E[\varphi]=i\tilde S_E[\varphi]$ in Euclidean signature to a real action $\Gamma_E[\varphi]$ on a macroscopic scale. Translated back to Minkowski signature, this implies that for strong interactions, an imaginary action $S[\phi]=i\tilde S[\varphi]$ corresponding to a positive weight
\begin{equation}
e^{iS[\varphi]} = e^{-\tilde S[\varphi]}
\label{eq:}
\end{equation}
could lead to an effective action $\Gamma[\varphi]$ of the usual form.

This opens the possibility for a probability interpretation of quantum field theory in terms of the bilocal action $Q[\chi]$. Before we discuss this in more detail in the following section, let us make some remarks on the connection of the formalism to usual quantum field theory. 

(1) For free theories where $\Gamma_k^{(3)}$ and higher terms vanish, the inverse propagator $\Gamma_k^{(2)}$ is not renormalized. This implies in particular that its complex phase remains constant so that the above mechanism is not applicable. Interaction effects are crucial in this respect.

(2) One might wonder, whether the proposed scenario can be applied to the Standard model, where most interaction parameters are rather weak. However, although one needs a strong renormalization group running, it is nevertheless possible that the infrared physics is dominated by weak coupling constants. The reason is that in most cases strong interactions are strongly screened by quantum fluctuations leading to effective theories with small interaction parameters in the infrared. It may be that the strong renormalization group running which drives the coefficients $\gamma_k^{(n)}$ towards the real axis happens at very large scales, possibly above all other relevant scales such as the electroweak scale or the Planck scale.

(3) We note that a given (real) effective action $\Gamma[\phi]$ on the macroscopic scale can follow from several microscopic actions. In particular, since the theory-subspace with $\alpha_k=0$ corresponds to a fixed point for $\alpha_k$, there will usually be a trajectory in this subspace leading to a particular real effective action $\Gamma[\phi]$ at $k=0$.

(4) Usually one assumes that the effective action $\Gamma[\varphi]$ in Minkowski space has to be real since this corresponds to unitarity of the S-matrix. Euclidean quantum field theories must have Osterwalder-Schrader reflection positivity which implies positivity of the norm for Minkowski signature \cite{OsterwalderSchrader}. It is clear that the scenario proposed here leads to deviations from positivity in the microscopic regime. To prevent conflicts with well established and experimentally tested features of quantum mechanics, unitarity or Osterwalder-Schrader positivity must be effectively restored at macroscopic scales. More detailed investigations for concrete models must reveal that this is actually the case.

(5) Finally, we remark that deviations from unitarity in the microscopic regime could lead to measureable effects that could be tested in experiments. Presumably this is not possible if unitarity is violated only at very large momentum scales but more detailed investigations are necessary to clarify this point, as well.

\section{Properties of bilocal statistical field theory}
\label{sec:Propertiesbilocal}
In this section we assume for the time being that a mechanism as the one proposed in the previous section allows to reformulate quantum field theory in terms of a bilocal statistical field theory. We investigate some consequences and discuss how known physics emerges. 

On the first sight, statistical field theory for bilocal fields $\chi(x,y)$ and $\chi^*(x,y)$ with a functional probability 
\begin{equation}
e^{-Q[\chi]}
\label{eq:lo1}
\end{equation}
is very different from the quantum mechanical description we are used to. For example, Eq.\ \eqref{eq:lo1} describes probabilities for field configurations in space-time, while quantum mechanics describes probabilities for measuring particles in a certain state conditional to some preparation experiment in the past. 
This is irritating if there is only one world or only one space-time so that such a probability does not seem to make much sense. However, one can interpret Eq.\ \eqref{eq:lo1} as describing probabilities for microscopic field configurations which cannot be directly accessed by measurements. We will argue below that only particular modes correspond to particles while others should be seen as part of a non-trivial vacuum. Non-Gaussian terms in $Q[\chi]$ lead to connections between these two sectors. It is natural to employ a statistical ensemble for the microscopic or vacuum part of $\chi$.

Most fluctuations of the bilocal field will decay on short time scales or the values of $\chi$ will be uncorrelated after some relatively short time. Such fluctuations correspond to the vacuum. A typical particle detector produces signals only if an excitation is coherent for a long enough time. For a statistical field theory the relevant object is the two-point correlation function. For a bilocal theory it reads
\begin{eqnarray}
\nonumber
&&G(x_+,x_-;y_+,y_-)\\ 
&&= \frac{1}{Z[K=0]} \int D \chi\; \chi(x_+,x_-) \chi^*(y_+,y_-).
\label{eq:}
\end{eqnarray}
The inverse propagator $G^{-1}$ is defined by
\begin{eqnarray}
\nonumber
&&\int_{z_+,z_-} G^{-1}(x_+,x_-;z_+,z_-) G(z_+,z_-;y_+,y_-) \\
&&= \delta(x_+-y_+) \delta(x_- -y_-).
\label{eq:}
\end{eqnarray}
Typically, it can be written as
\begin{equation}
G^{-1} (x_+,x_-;y_+,y_-) = \delta(x_+-y_+) \delta(x_--y_-) P_+ P_-^\dagger,
\label{eq:}
\end{equation}
where $P_+$ and $P_-$ contain derivatives with respect to $x_+$ and $x_-$. Modes of $\chi$ that correspond to long range fluctuations are characterized by
\begin{equation}
P_+ P_-^\dagger \chi(x,y) = 0.
\label{eq:nund1}
\end{equation}
Short range fluctuations have a certain influence on the form of $G^{-1}$, in particular they change the values of appearing constants such as the mass or coefficients of derivative terms. This are the well known renormalization group modifications.

Equation \eqref{eq:nund1} cannot be a sufficient condition for a mode of $\chi$ to be directly measurable in experiments. The field $\chi$ still has two independent variables $x$ and $y$. This is in contrast to our experience of a ``local world'' where one can only measure fields depending on a single space-time coordinate.

One can make progress by assuming that a mode that is detectable needs not only long-range (or long-time) correlations but also positive ones. By this we mean that the two-point correlation function $G$ needs to be real and positive semi-definite. Modes that have this property are the ``positive correlated modes'' discussed in section \ref{sec:positiviCorrelations}. To discuss the measurement of particles in a particular basis $\phi_n(\vec x)$ it is useful to write the field $\chi$ at equal time argument as
\begin{eqnarray}
\label{eq:nund2}
\chi((t,\vec x),(t,\vec y)) &=& \sum_n \chi((t,n),(t,n))\; \phi_n(\vec x) \phi_n^*(\vec y)\\
\nonumber
&&+ \sum_{n\neq m} \chi((t,n),(t,m))\; \phi_n(\vec x) \phi_m^*(\vec y).
\end{eqnarray}
The first sum on the right hand side describes the contribution of positive modes while the second sum adds off-diagonal terms. The diagonal terms are the ones that describe particles according to the rules of quantum mechanics. We emphasize that the fact that the correlations of $\chi((t,n),(t,n))$ are real and positive semi-definite is a special property of the functional probabilities $e^{-Q[\chi]}$ that can be written as a double functional integral as in Eq.\ \eqref{eq:defQ}. For more general bilocal theories this will not be the case. 

We note that this can explain why our world is ``to large extend local'' although the field $\chi$ is bilocal. The ``positive correlated modes'' $\chi((t,n),(t,n))$ have only one free time argument remaining. The spatial dependence is encoded by a function $\phi_n(\vec x)$ as in Eq.\ \eqref{eq:lo2}. In many cases this function will be localized in some volume of space and we encounter the same kind (and degree) of nonlocality as in quantum mechanics. 

A useful physical picture is that the fields that are relevant for experiments are coarse-grained or averaged over some time interval $\Delta t$. The coarse-grained fields might be given by
\begin{eqnarray}
\label{eq:nund3}
\chi_r((t,\vec x),(t,\vec y)) &=& A\; \chi ((t,\vec x),(t,\vec y))\\
\nonumber
&=& \frac{1}{\Delta t} \int_{t^\prime} a\left(\frac{t-t^\prime}{\Delta t}\right) \; \chi((t^\prime,\vec x),(t^\prime,\vec y)),
\end{eqnarray}
where $a(x)$ is a positive function that vanishes for $|x|\gg 1$ and has the property $\int a(x) d x=1$. If a mode of $\chi$ is oscillating with frequency large compared to $1/\Delta t$ it will be effectively annihilated by the operator $A$. Oscillating modes typically correspond to negative or more general complex two-point correlation function. In contrast, modes where $G$ is real and positive are usually constant and therefore non annihilated by the coarse-graining. Note however that also the diagonal part of $\chi$, i.e. the first term on the right hand side of Eq.\ \eqref{eq:nund2}, might contain terms that are annihilated by the coarse-graining in a concrete situation.

Let us now discuss some physical situations to find out whether the picture that emerged so far makes sense. We first consider a situation where a particle has been prepared in an eigenstate of the (single particle) Hamiltonian labeled by $n=1$ at time $t$. This situation corresponds to the coarse-grained fields
\begin{eqnarray}
 \nonumber
\chi((t,1),(t,1)) &\neq& 0,\\
\chi((t,n),(t,n)) &=& 0 \; \text{for}\; n\neq 1.
\end{eqnarray}
The fields at later times $t^\prime>t$ are positive correlated to the ones at time $t$ and since $n$ labels eigenstates of the Hamiltonian, the cross-correlations vanish. In other words we have
\begin{equation}
 \langle \chi((t^\prime, n^\prime),(t^\prime, n^\prime)) \chi^*((t,n),(t,n)) \rangle = \text{const}\; \delta_{n^\prime n}.
\end{equation}
This implies that only the mode with $n=1$ will be nonzero at later times. This situation is precisely as we expect it form the laws of quantum mechanics.

Now let us consider the situation where the measurement at time $t^\prime$ is performed in another basis which we label by the index $\tilde n$. We have now correlation functions
\begin{equation}
 \langle \chi((t^\prime, \tilde n^\prime),(t^\prime, \tilde n^\prime)) \chi^*((t,n),(t,n)) \rangle = \text{const} \; p_{\tilde n^\prime n}
\label{eq:lnu1}
\end{equation}
that equal by construction the transition probabilities from quantum mechanics $p_{\tilde n^\prime n}$ (up to a constant factor). 

Similarly, one can expect that other correlation functions are equal for quantum field theory and our bilocal statistical field theory. However, when one attempts to make this more precise, one is confronted with a puzzle: Why does a particle-like excitation not split into several ones? Indeed, one could expect nonzero field $\chi((t^\prime,\tilde n),(t^\prime, \tilde n))$ for several $\tilde n$ if the correlation function in Eq.\ \eqref{eq:lnu1} with $n=1$ is nonzero. The question is closely connected with the question why particle number is conserved and in particular why it is discrete. For a theory with a global $U(1)$ symmetry one can derive conservation laws for appropriate currents. However, this does not imply the discreteness of the particle number. It may well be that this puzzle can be solved once the formalism is extended to fermionic theories. A certain kind of discreteness arises naturally for functional integrals over Grassmann numbers. However, this is beyond the scope of the present paper and we leave this point open for future investigations.

To close this section we remark that the basis chosen to perform the initial preparation at time $t$ does not have to be the eigenbasis of the Hamiltonian. It is straightforward to transfer the above considerations to more general situations.

\section{Conclusions}
\label{sec:conclusions}
We have developed a formulation of quantum field theory for bosonic scalar fields in terms of a functional integral over bilocal fields. This has the advantage that physical observables such as correlation functions or transition probabilities can be obtained directly instead of first calculating probability amplitudes which need to be squared to give probabilities. 

The bilocal formulation can be interpreted in terms of a statistical field theory if the bilocal action $Q[\chi]$ is real. An explicit example for this corresponds to a quantum field theory where unitarity is broken at microscopic scales. The flow equation for the effective average action in Euclidean space suggests that is becomes real at macroscopic scales even if the microscopic action is imaginary. Strong fluctuation effects can lead to an effective action of the usual form at small values of the cutoff parameter. In Minkowski space this corresponds to a theory where the weight of the functional integral is real and positive. This implies in particular that the exponentiated bilocal action $e^{-Q[\chi]}$ is positive. 
Unitarity would be an emerging phenomenon instead of being a property of the fundamental theory. 

It is clear that this scenario needs (and deserves) further investigations. A more detailed renormalization group analysis of statistical field theories with Minkowski signature could reveal interesting relations to corresponding quantum field theories. By investigating concrete and realistic models one could explore the experimental signatures for deviations from unitarity in the microscopic regime

We have also discussed how properties of particles can arise in a bilocal statistical field theory. Positive correlations at subsequent times play an important role. Quantum mechanical observables can be represented as correlation functions of bilocal fields. 

It is planed to extend the formalism to fermionic theories described by Grassmann valued fields. The bilocal field will always be of bosonic nature since it corresponds to a pair of original fields, but one can expect that some properties will differ. The formalism in its present form already provides evidence for the intriguing possibility that quantum field theory can be understood as a bilocal statistical field theory.

\end{document}